%% file: albright_proc7.tex
\documentclass[
  ,final            
  ]
  {aipproc}

\layoutstyle{8x11single}

\usepackage[latin1]{inputenc}
\usepackage[T1]{fontenc}
\usepackage[english]{babel}
\usepackage[pdftex]{xcolor}
\usepackage{graphicx}
\usepackage{xspace,tabularx}

\newcommand\order[1]{\ensuremath{\mathcal{O}(#1)}}

\newcommand{\SU}[1]{\ensuremath{{\rm SU}(#1)}}
\newcommand{\U}[1]{\ensuremath{{\rm U}(#1)}}
\newcommand{\SO}[1]{\ensuremath{{\rm SO}(#1)}}

\newcommand{\irrep}[1]{\ensuremath{\boldirrep{#1}}}

\newcommand{\boldirrep}{\mathbf}
\newlength{\irrepwidth}
\newlength{\irrepbarthickness}
\newlength{\irrepbarheight}
\newcommand{\irrepbar}[1]{%
	\settoheight{\irrepbarheight}{\ensuremath{\boldirrep{#1}}}%
	\settowidth{\irrepwidth}{\ensuremath{\boldirrep{#1}}}%
	\makebox[0pt][l]{\ensuremath{\boldirrep{#1}}}%
	\rule[1.2\irrepbarheight]{\irrepwidth}{\irrepbarthickness}%
}
\newcommand{\irrepvar}[1]{(\irrep{#1})}
\newcommand{\irrepbarvar}[1]{(\irrepbar{#1})}

\newcommand{\irrepsub}[2]{\ensuremath{\irrep{#1}_{#2}}}
\newcommand{\irrepbarsub}[2]{\ensuremath{\irrepbar{#1}_{#2}}}

\newcommand\fermion\irrepsub
\newcommand\fermionbar\irrepbarsub

\newcommand{\higgs}[2]{\ensuremath{\irrep{#1}_{\rm H}}}
\newcommand{\higgsbar}[2]{\ensuremath{\irrepbar{#1}_{\rm H}}}

\newcommand{\massivefermionpair}[2]{%
\ensuremath{#1{\times}#2}%
}

\newcommand{\ldot}{{.}}

\begin{document}

\begin{flushright}
  CETUP*-12/013\\
  FERMILAB-CONF-12-527-T\\
\end{flushright}
\skip -0.5in
\title{An Explicit \SU{12} Family and Flavor Unification Model}


\author{Carl H. Albright}{
  address={Department of Physics, Northern Illinois University, DeKalb, IL 60115}, 
  altaddress={Theoretical Physics, Fermilab, Batavia, IL 60510}
}

\author{Robert P. Feger}{
  address={Department of Physics and Astronomy, Vanderbilt University, Nashville, TN 37235}
}

\author{Thomas W. Kephart}{
  address={Department of Physics and Astronomy, Vanderbilt University, Nashville, TN 37235}
}

\begin{abstract}
An explicit SUSY SU(12) unification model with three light chiral families is presented
which avoids any external flavor symmetries.  The hierarchy of quark and lepton 
masses and mixings is explained by higher dimensional Yukawa interactions involving Higgs 
bosons containing \SU{5} singlet fields with VEVs appearing at or below the SUSY GUT scale
of $2 \times 10^{16}$ GeV, approximately 50 times smaller than the \SU{12} unification scale.  
The model has been found to be in good agreement with the observed quark and lepton 
masses and mixings, with nearly all prefactors of $\mathcal{O}(1)$ in the four Dirac 
and one Majorana fermion mass matrices.
\end{abstract}

\maketitle  


\section{Motivation}

Grand Unified Theories (GUTs) of quarks and leptons have played an 
important role for almost 40 years in theoretical attempts to 
make sense of the apparent group structure and mass spectra of  
the quark and lepton matter fields and coupling strengths of the 
gauge and Higgs fields observed in Nature.  Although many varieties
of models have been constructed, the most popular unified ones are 
based on the unitary, orthogonal, or exceptional groups \SU{5},
\SO{10}, or E$_6$, respectively.  But in each of these cases, the 
chiral irreducible representations (irreps) can uniquely describe 
only one family of quarks and leptons: towit, ${\bf 10,\ \overline{5}}$,
and ${\bf 1}$ for \SU{5}, ${\bf 16}$ for \SO{10}, and ${\bf 27}$
for E$_6$.  In order to accommodate the three families observed todate, 
it has been conventional to introduce in addition to one of the 
above $G_{\rm family}$ groups, a $G_{flavor}$ symmetry group which
also distinguishes the families.  While continuous flavor symmetries
such as \U{1}, \SU{2}, \SU{3} and their products have been considered
in the past, more recently discrete symmetry groups such as $A_4,\ 
T'$ and $S_4$, etc. have been fashionable in the past 10 years 
\cite{af,ikoso}.  In either case, the GUT model then involves the 
direct product group $G_{family} \times G_{flavor}$.

True family and flavor unification requires the introduction of a 
higher rank simple group.  Some earlier studies along this line have
been based on \SO{18}~\cite{GellMann:1980vs,Fujimoto:1981bv}, 
\SU{11}~\cite{Georgi:1979md,Kim:1981bb}, and 
\SU{9}~\cite{Frampton:1979cw,Frampton:1979fd}.  More
recently, models based on \SU{7}~\cite{Barr:2008gz}, 
\SU{8}~\cite{Barr:2008pn}, and \SU{9} again~\cite{Frampton:2009ce,Dent:2009pd},
the latter reference by two of us (RF and TWK),  
but none have been totally satisfactory due to a huge number of unwanted 
states and/or unsatisfactory mass matrices.  Here we describe an \SU{12} 
unification model~\cite{afksu12} with interesting features that was constructed
with the help of a Mathematica computer package called LieART 
written by two of us (RPF and TWK)~\cite{fgLieART}.  
This program allows one to compute tensor products, branching rules, etc., 
and perform detailed searches for satisfactory models in a timely fashion.
While other smaller and larger rank unitary groups were examined, 
a model based on \SU{12} appeared to be the most satisfactory 
minimal one for our purpose.  We sketch here the model construction and
point out that further details can be found in ~\cite{afksu12}.

\section{\SU{12} Unification Model and Particle Assignments}

While the three popular GUT groups cited earlier each have just 
one useful chiral irreducible representation (irrep), \SU{12} has 
11 totally antisymmetric irreps: 
$\irrep{12},\ \irrep{66},\ \irrep{220},\ \irrep{495},
\ \irrep{792},\ \irrep{924},\ \irrepbar{792},
\ \irrepbar{495},\ \irrepbar{220},\ \irrepbar{66}$,
and $\irrepbar{12}$, of which 10 are complex (while \irrep{924} is real),
which allow three \SU{5} families to be assigned to different chiral irreps.  
For this purpose, one chooses an anomaly-free set of \SU{12} irreps
which contains three chiral \SU{5} families and pairs of fermions
which will become massive at the \SU{5} scale.  One such suitable
set consists of 
\begin{equation}\label{eq:AnomFreeSet}
    6\irrepvar{495} + 4\irrepbarvar{792} + 4\irrepbarvar{220} +
    \irrepbarvar{66} + 4\irrepbarvar{12} 
    \rightarrow 3(\irrep{10} + \irrepbar{5} + \irrep{1}) + 
    238(\irrep{5} + \irrepbar{5}) + 211(\irrep{10} + \irrepbar{10})
    + 484(\irrep{1}) 
\end{equation}

\noindent where the decomposition to anomaly-free \SU{5} states
has been indicated. The latter follows from the \SU{12} $\rightarrow$
\SU{5} branching rules:
\begin{eqnarray}\label{eq:branchrules}
    \irrep{495} &\rightarrow & 35\irrepvar{5} + 21\irrepvar{10} + 
      7\irrepbarvar{10} + \irrepbar{5} + 35\irrepvar{1},\\
    \irrepbar{792} &\rightarrow& 7\irrepvar{5} + 21\irrepvar{10} + 
      35\irrepbarvar{10} + 35\irrepbarvar{5} + 22\irrepvar{1},\\
    \irrepbar{220} &\rightarrow& \irrep{10} + 7\irrepbarvar{10}
      + 21\irrepbarvar{5} + 35\irrepvar{1},\\
    \irrepbar{66} &\rightarrow& \irrepbar{10} + 7\irrepbarvar{5} 
      + 21\irrepvar{1},\\
    \irrepbar{12} &\rightarrow& \irrepbar{5} + 7\irrepvar{1}
\end{eqnarray}

A search through the possible assignments of the three light chiral 
families to the \SU{12} irreps appearing in the anomaly-free set
of Eq. (1) reveals the following selection for a satisfactory low
scale phenomenology:
\begin{equation}\label{eq:famassign}
\begin{array}{lrcl}
  {\rm 1st\ Family:} & \irrepvar{10}\irrep{495_1} & \supset & u_L,\ u^c_L,\ d_L,\ e^c_L\\
              & \irrepbarvar{5}\irrepbar{66_1} & \supset & d^c_L,\ e_L,\ \nu_{1L}\\
	      & \irrepvar{1}\irrepbar{792_1} & \supset & N^c_{1L}\\
  {\rm 2nd\ Family:} & \irrepvar{10}\irrepbar{792_2} & \supset & c_L,\ c^c_L,\ s_L,\ \mu^c_L\\
              & \irrepbarvar{5}\irrepbar{792_2} & \supset & s^c_L,\ \mu_L,\ \nu_{2L}\\
              & \irrepvar{1}\irrepbar{220_2} & \supset & N^c_{2L}\\
  {\rm 3rd\ Family:} & \irrepvar{10}\irrepbar{220_3} & \supset & t_L,\ t^c_L,\ b_L,\ \tau^c_L\\
              & \irrepbarvar{5}\irrepbar{792_3} & \supset & b^c_L,\ \tau_L,\ \nu_{3L}\\
              & \irrepvar{1}\irrepbar{12_3} & \supset & N^c_{3L}
\end{array}
\end{equation}

\noindent Here the subscripts on the \SU{12} irreps refer to the family in question, 
while the numbers in parentheses are just the \SU{5} irreps chosen.  Note that each 
\SU{5} family multiplet can be uniquely assigned to a different \SU{12} multiplet
in the anomaly-free set according to (1).  On the other hand, the remaining \SU{5}
multiplets are unassigned but form conjugate pairs which  become massive and decouple 
at the \SU{5} scale and are of no further interest.

\section{Effective Theory Approach and Leading Order Tree Diagrams}

We start with the \SU{12} model sketched above and take it to be supersymmetric.  With 
a $\irrep{143_H}$ adjoint Higgs field present, the breaking of \SU{12} to \SU{5} can
occur via \SU{12} $\rightarrow$ \SU{5} $\times$ \SU{7} $\times$ \U{1}, and in steps down to 
\SU{5} via a set of antisymmetric chiral superfield irreps appropriately chosen to 
preserve supersymmety \cite{pFtwK1,pFtwK2}.  Unbroken supersymmetry at the \SU{5} GUT 
scale allows us to deal only with tree diagrams in order to generate higher 
dimensional operators, for loop corrections are much suppressed.  

For this purpose, we introduce massive $\irrep{220} \times \irrepbar{220}$ and 
$\irrep{792} \times \irrepbar{792}$ fermion pairs at the \SU{12} scale.  In addition,
we introduce $\irrepvar{1}\irrep{66_H},\ \irrepvar{1}\irrepbar{66_H}$, and 
$\irrepvar{1}\irrep{220_H},\ \irrepvar{1}\irrepbar{220_H}$ conjugate Higgs pairs which
acquire \SU{5} singlet VEVs at the SUSY \SU{5} GUT scale.   Finally, doublets in 
$\irrepvar{5}\irrep{924_H}$ and $\irrepbarvar{5}\irrep{924_H}$ Higgs fields effect the 
electroweak symmetry breaking at the electroweak scale.  The list comprises then the 
following:

\begin{equation}
    \begin{array}{llc}
        \multicolumn{2}{c}{\rm Higgs\ Bosons}  &  {\rm Massive\ Fermions}\\
        (\irrep{5})\higgs{924},& (\irrepbar{5})\higgs{924}, & \irrep{220}{\times} \irrepbar{220},\\
        (\irrep{1})\higgs{66}, & (\irrep{1})\higgsbar{66},   & \irrep{792}{\times} \irrepbar{792}\\
        (\irrep{1})\higgs{220},& (\irrep{1})\higgsbar{220},  & \\
        (\irrep{24})\higgs{143} &\\
    \end{array}
\end{equation}

\noindent  For each element of the quark and lepton mass matrices, tree diagrams can then be
constructed from three-point vertices which respect the \SU{12} and \SU{5} multiplication rules.

For illustration we present the lowest order tree diagram contributions to the 33 elements 
for the up and down quark mass matrices, taking into account the family assignments in 
\eqref{eq:famassign}. These are listed as ${\bf U33}$ and ${\bf D33}$, respectively in \eqref{eq:figs}. 
The convention is adopted that the left-handed fields appear on the left and the left-handed 
conjugate fields appear on the right.  
\begin{equation}
\label{eq:figs}
   \textbf{U33:}\quad\raisebox{-.38\height}{\includegraphics[scale=0.85]{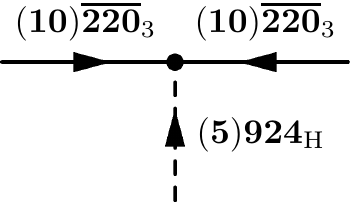}}\qquad\qquad
    \textbf{D33:}\quad\raisebox{-.45\height}{\includegraphics[scale=0.85]{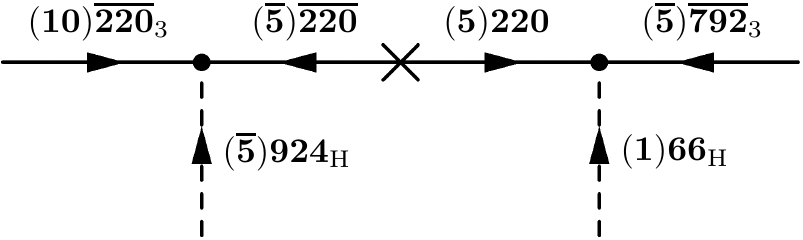}}
\end{equation}
\noindent For convenience we introduce the following short-hand notation to describe each of these diagrams:
\begin{equation}\label{eq:33diagrams}
\begin{array}{llcll}
    {\bf U33}: & \irrepvar{10}\irrepbar{220_3}.\irrepvar{5}\irrep{924_H}.\irrepvar{10}
             \irrepbar{220_3},   & \qquad &   {\bf D33}: & 
           \irrepvar{10}\irrepbar{220_3}.\irrepbarvar{5}\irrep{924_H}.\irrepbarvar{5}\irrepbar{220}
	     \times \irrepvar{5}\irrep{220}.\irrepvar{1}\irrep{66_H}.\irrepbarvar{5}
             \irrepbar{792_3},\\
\end{array}
\end{equation} 

\noindent  The leading order term for {\bf U33} is seen to have dim-4, while that for {\bf D33} has dim-5, 
due to the $\irrepvar{1}\irrep{66_H}$ \SU{5} Higgs singlet insertion resulting in one extra external
Higgs field. 

The full sets of leading order up- and down-quark diagrams for each matrix element is presented in Table I,
while those for the Dirac and Majorana neutrino diagrams are listed in Table II.  It is rather remarkable 
that only one diagram for each matrix element appears at leading order for all four mass matrices.
\input{QuarkMassTermDiagramsTablec.tex}

\section{Mass Matrices and Mixings}

Given the leading-order diagrams for each matrix element in Tables 1 and 2, we can 
then construct the quark and lepton mass matrices as follows.  To each diagram 
corresponds a coupling constant or prefactor, $h^u_{ij},\ h^d_{ij},\ h^{dn}_{ij}$ or 
$h^{mn}_{ij}$ for the $ij$th element of the appropriate mass matrix, which is assumed to be
of order one at the \SU{12} unification scale, as naturalness predicts.    
Every \SU{5} Higgs singlet insertion in higher-order tree diagrams introduces one power of 
$\varepsilon \equiv M_{\SU{5}}/M_{\SU{12}} \sim 1/50$ through the appearance of the ratio of the 
singlet Higgs VEV to the mass of the conjugate fermion fields after the latter are integrated out.
Finally as a result of the electroweak spontaneous symmetry breaking, the ${\bf 924_H}$ acquires a 
weak scale VEV, ${\rm v}$.  Hence for the two quark diagrams illustrated, the matrix element contributions 
are
\begin{equation}
\label{eq:33contributions}
     {\bf U33}: h^u_{33}{\rm v}\ t^T_Lt^c_L, \hspace{1in} {\bf D33}: h^d_{33}\epsilon {\rm v}\ b^T_L b^c_L. \\
\end{equation}

\noindent Note that as a result of the chiral \SU{5} irrep structure, the lowest order tree 
diagram contribution to the 33 element of the charged lepton mass matrix is just the reflection 
of the diagram for the down quark 33 mass matrix element about the center of the diagram.  
Thus its 33 matrix element contribution is just the transpose of {\bf D33}.  More generally, 
the prefactors are related by $h^\ell_{ij} = h^d_{ji}$.  By the same reasoning, it is clear that 
the up quark mass matrix elements are symmetric under interchange of $i$ and $j$.

\input{NeutrinoMassTermDiagramsTablec.tex}

From Table 1 we then see that the two quark and charged lepton mass matrices are given by 
\begin{equation}
    \begin{array}{rl}
        M_U =&
        \left(\matrix{
            h^u_{11}\epsilon^4 & h^u_{12}\epsilon^3  & h^u_{13}\epsilon^2  \\
            h^u_{12}\epsilon^3 & h^u_{22}\epsilon^2  & h^u_{23}\epsilon    \\
            h^u_{13}\epsilon^2 & h^u_{23}\epsilon    & h^u_{33}        \\}\right) \!{\rm v}\:,\\[0.3in]
        M_D =&
        \left(\matrix{
            h^d_{11}\epsilon^4 & h^d_{12}\epsilon^3  & h^d_{13}\epsilon^3  \\
            h^d_{21}\epsilon^3 & h^d_{22}\epsilon^2  & h^d_{23}\epsilon^2  \\
            h^d_{31}\epsilon^2 & h^d_{32}\epsilon    & h^d_{33}\epsilon    \\}\right) \!{\rm v}\:,\\[0.3in]
        M_L =&
        \left(\matrix{
            h^\ell_{11}\epsilon^4 & h^\ell_{12}\epsilon^3  & h^\ell_{13}\epsilon^2  \\
            h^\ell_{21}\epsilon^3 & h^\ell_{22}\epsilon^2  & h^\ell_{23}\epsilon    \\
            h^\ell_{31}\epsilon^3 & h^\ell_{32}\epsilon^2  & h^\ell_{33}\epsilon    \\} \right)\!{\rm v} 
            = M_D^T.
    \end{array}
\end{equation}

\noindent  While the up-quark matrix is symmetric, the
down-quark and charged-lepton mass matrices are doubly lopsided in that the terms with
$h^d_{23}$ and $h^l_{32}$ are suppressed by one extra power of $\epsilon$
compared with the $h^d_{32}$ and $h^l_{23}$ terms, respectively. For $M_D$,
for example, this implies that a larger right-handed rotation than
left-handed rotation is needed to bring the down quark matrix into diagonal
form, while the opposite is true for $M_L$.

With the heavy right-handed neutrinos assigned to \SU{5} singlets in \eqref{eq:famassign}, 
the resulting Dirac and Majorana neutrino 33 mass matrix elements receive the 
following dim-4 contributions as seen from Table 2: 
\begin{equation}
\label{eq:neutrinocont}
\begin{array}{rl}
      {\bf DN33}:  h^{dn}_{33} {\rm v} \overline{\nu}_{3L}N^c_{3L}, \qquad & \qquad 
         {\bf MN33}: h^{mn}_{33} \Lambda_R  N^{c^T}_{3L} N^c_{3L}.\\
\end{array}
\end{equation}

\noindent Here $\Lambda_R$ 
represents the right-handed mass scale, typically of $\order{10^{14}}$ GeV, whereas
the \SU{5} SUSY GUT scale is $2 \times 10^{16}$ GeV to obtain gauge coupling unification.  
Again, a factor of $\epsilon$ enters for every singlet Higgs insertion in higher order diagrams.  
The two neutrino mass matrices can then be read off from Table 2, and we find 
\begin{equation}
    \begin{array}{rl}
        M_{DN} =&
        \left(\matrix{
            h^{dn}_{11}\epsilon^3 & h^{dn}_{12}\epsilon^2 & h^{dn}_{13}\epsilon \\
            h^{dn}_{21}\epsilon^2 & h^{dn}_{22}\epsilon   & h^{dn}_{23} \\
            h^{dn}_{31}\epsilon^2 & h^{dn}_{32}\epsilon   & h^{dn}_{33} \\}\right)\!{\rm v}\:,\\[0.3in]
        M_{MN} =&
        \left(\matrix{
            h^{mn}_{11}           & h^{mn}_{12}\epsilon   & h^{mn}_{13}\epsilon^2 \\
            h^{mn}_{12}\epsilon   & h^{mn}_{22}\epsilon^2 & h^{mn}_{23}\epsilon^3 \\
            h^{mn}_{13}\epsilon^2 & h^{mn}_{23}\epsilon^3 & h^{mn}_{33}  \\}\right)\!\Lambda_R.\\
    \end{array}
\end{equation}

\noindent where ${M_{DN}}$ is also double lopsided, while ${M_{MN}}$ is 
complex symmetric as usual.  The symmetric light-neutrino mass matrix is obtained 
via the Type I seesaw mechanism:
\begin{equation}
    M_\nu = -M_{\rm DN}M_{\rm MN}^{-1}M_{\rm DN}^T.
\end{equation}
Keeping only the leading-order terms in $\epsilon$ for each matrix element, we find
\begin{equation}
    M_\nu \approx \frac{{\rm v}^2}{\Lambda_R}\times\!\!\\
        \left(\begin{array}{ccc}
            \epsilon ^2 \left(\frac{h^{dn^2}_{12} h^{mn}_{11}}{h^{mn^2}_{12}{-}h^{mn}_{11} h^{mn}_{22}}{-}\frac{h^{dn^2}_{13}}{h^{mn}_{33}}\right)
            & \epsilon  \left(\frac{h^{dn}_{12} h^{dn}_{22} h^{mn}_{11}}{h^{mn^2}_{12}{-}h^{mn}_{11} h^{mn}_{22}}{-}\frac{h^{dn}_{3} h^{dn}_{23}}{h^{mn}_{33}}\right)
            & \epsilon  \left(\frac{h^{dn}_{12} h^{dn}_{32} h^{mn}_{11}}{h^{mn^2}_{12}{-}h^{mn}_{11} h^{mn}_{22}}{-}\frac{h^{dn}_{13} h^{dn}_{33}}{h^{mn}_{33}}\right)\\[0.1in]
            \epsilon  \left(\frac{h^{dn}_{12} h^{dn}_{22} h^{mn}_{11}}{h^{mn^2}_{12}{-}h^{mn}_{11} h^{mn}_{22}}{-}\frac{h^{dn}_{13} h^{dn}_{23}}{h^{mn}_{33}}\right)
            & \frac{h^{dn^2}_{22} h^{mn}_{11}}{h^{mn^2}_{12}{-}h^{mn}_{11} h^{mn}_{22}}{-}\frac{h^{dn^2}_{23}}{h^{mn}_{33}}
            & \frac{h^{dn}_{22} h^{dn}_{32} h^{mn}_{11}}{h^{mn^2}_{12}{-}h^{mn}_{11} h^{mn}_{22}}{-}\frac{h^{dn}_{23} h^{dn}_{33}}{h^{mn}_{33}} \\[0.1in]
            \epsilon  \left(\frac{h^{dn}_{12} h^{dn}_{32} h^{mn}_{11}}{h^{mn^2}_{12}{-}h^{mn}_{11} h^{mn}_{22}}{-}\frac{h^{dn}_{13} h^{dn}_{33}}{h^{mn}_{33}}\right)
            & \frac{h^{dn}_{22} h^{dn}_{32} h^{mn}_{11}}{h^{mn^2}_{12}{-}h^{mn}_{11} h^{mn}_{22}}{-}\frac{h^{dn}_{23} h^{dn}_{33}}{h^{mn}_{33}}
            & \frac{h^{dn^2}_{32} h^{mn}_{11}}{h^{mn^2}_{12}{-}h^{mn}_{11} h^{mn}_{22}}{-}\frac{h^{dn^2}_{33}}{h^{mn}_{33}}\\  \end{array} \right)
\end{equation}
which does not involve the prefactors $h^{dn}_{11}$, $h^{dn}_{21}$, $h^{dn}_{31}$,
$h^{mn}_{13}$ and $h^{mn}_{23}$.
  
The light-neutrino mass matrix exhibits a much milder hierarchy compared to
the up-type and down-type mass matrices, as can be seen from the pattern of
powers of $\epsilon$. A mild or flat hierarchy of $M_\nu$ is conducive to
obtaining large mixing angles and similar light neutrino masses. Furthermore,
one observes that the light neutrino mass matrix obtained via the
seesaw mechanism involves the doubly lopsided Dirac neutrino mass matrix twice. The
lopsided feature of $M_{DN}$ is such as to require a large left-handed
rotation to bring $M_\nu$ into diagonal form.

\section{Numerical Results}

From the above up and down quark, charged lepton and light neutrino mass matrices, 
one can diagonalize the corresponding Hermitian matrices, $MM^\dagger$, in the usual 
manner to obtain the mass eigenvalues and the unitary transformations, $U$, effecting
the diagonalizations.  The Cabbibo-Kobayashi-Maskawa (CKM) quark mixing matrix
and the corresponding Pontecorvo-Maki-Nakagawa-Sakata (PMNS) lepton mixing matrix
then follow as 
\begin{equation}
\begin{array}{rl}
   V_{CKM} = U^\dagger_U U_D \qquad\qquad V_{PMNS} = U^\dagger_L U_\nu
\end{array}
\end{equation}

To obtain numerical results for the model predictions, we evaluate the mass matrices
at the top quark mass scale and use just real prefactors to avoid too many fit 
parameters for good fit convergence.  There are 6 prefactors each for the symmetric up
quark and Majorana matrices, 9 each for the lopsided down quark and Dirac neutrino
matrices, but 5 of them do not appear in the light neutrino mass matrix, making a 
total of 25 parameters.  In addition, we have one for the right-handed neutrino scale,
$\Lambda_R$ plus a value for $\epsilon$ which we fix at $\epsilon = 1/6.5^2 = 0.0237$
again for good fit convergence, for a grand total of 26 adjustable fit parameters.
To avoid correlated data parameters, we make use of the 9 quark and charged lepton 
masses plus the 3 neutrino $\Delta m^2$'s and the 18 CKM and PMNS mixing parameters
taken to be real, for a total of 30 data points.  We refer the reader to our published
paper ~\cite{afksu12} 
for full details of the fitting procedure, where we have included 
the latest best value for the reactor neutrino mixing angle, $\theta_{13}$.  There 
can be found a table giving the phenomenological mass and mixing data entering the 
fit, as well as the theoretical mass and mixing results obtained from the fitting
procedure.

The best fit was obtained with a normal neutrino mass hierarchy with 
$\Lambda_R = 7.4 \times 10^{14}$ GeV and the following mass matrices:
\begin{equation}
\label{eq:fitmassmatrices}
    \begin{array}{rlrl}
        M_U =&
        \left(\matrix{
            -1.1\epsilon^4 & 7.1\epsilon^3  & 5.6\epsilon^2  \\
            7.1\epsilon^3 & -6.2\epsilon^2  & -0.10\epsilon    \\
            5.6\epsilon^2 & -0.10\epsilon    & -0.95        \\}\right) \!{\rm v}\:,\qquad\qquad
	    & M_D =&
        \left(\matrix{
            -6.3\epsilon^4 & 8.0\epsilon^3  & -1.9\epsilon^3  \\
            -4.5\epsilon^3 & 0.38\epsilon^2  & -1.3\epsilon^2  \\
            0.88\epsilon^2 & -0.23\epsilon    & -0.51\epsilon    \\}\right) \!{\rm v}\: = M^T_L,
	    \\[0.3in]
        M_{DN} =&
        \left(\matrix{
            h^{dn}_{11}\epsilon^3 & 0.21\epsilon^2 & -2.7\epsilon \\
            h^{dn}_{21}\epsilon^2 & -0.28\epsilon   & -0.15 \\
            h^{dn}_{31}\epsilon^2 & 2.1\epsilon   & 0.086 \\}\right)\!{\rm v}\:,\qquad\qquad
	    & M_{MN} =&
        \left(\matrix{
            -0.72           & -1.5\epsilon   & h^{mn}_{13}\epsilon^2 \\
            -1.5\epsilon   & 0.95\epsilon^2 & h^{mn}_{23}\epsilon^3 \\
            h^{mn}_{13}\epsilon^2 & h^{mn}_{23}\epsilon^3 & 0.093  \\}\right)\!\Lambda_R,\\[0.3in]
	M_\nu =&
	\left(\matrix{
	    -81\epsilon^2 & -4.3\epsilon & 2.4\epsilon \\
	    -4.3\epsilon  & -0.25        & 0.28 \\
	    2.4\epsilon   & 0.28 & -1.1 \\}\right)\ {\rm v}^2/{\Lambda_R}.\\[0.3in]
    \end{array}
\end{equation}

\noindent  Note that all prefactors except three in the above matrices are within a factor of 0.1 
to 10 of unity for this best fit.  The five independent prefactors, $h^{dn}_{11},\ h^{dn}_{21},\ 
h^{dn}_{31},\ h^{mn}_{13}$ and $h^{mn}_{23}$, do not influence the fit and 
remain undetermined as noted earlier.  For this best fit we find the neutrino mass values 
\begin{equation}
      m_1 = 0,\quad m_2 = 8.65,\quad m_3 = 49.7 {\rm \ meV};\qquad M_1 = 1.67 \times 10^{12},
\quad M_2 = 6.85 \times 10^{13},\quad M_3 = 5.30 \times 10^{14}\ {\rm GeV}.
\end{equation}

\noindent  In addition, the best fit favors $\delta_{CP} = \pi$ for the leptonic CP Dirac phase.
The value of $\epsilon$ used then implies that the \SU{12} GUT scale is about $M_{\SU{5}}/\epsilon
= 8.4 \times 10^{17}$ GeV, just below the reduced Planck scale, where we have used $2 \times 
10^{16}$ GeV for the \SU{5} unification scale.

All remaining mass and mixing parameters are fit quite well by the model; however, since $M_L$ is 
just the transpose of $M_D$ in leading order in $\epsilon$, the Georgi-Jarlskog relations~\cite{gj}
are not satisfied for the down quarks and charged leptons.  We have checked that the addition of an 
adjoint \higgs{143}, Higgs field whose VEV points in the $B - L$ direction contributes 
to $M_D$ and $M_L$ at one higher order of $\epsilon$, so that the $M_L = M^T_D$ relation is broken,
and more accurate values can be obtained for the down quark and charged lepton mass eigenvalues.

\section{Summary}

A unified \SU{12} SUSY GUT model was obtained by a brute force computer scan over many \SU{12} 
anomaly-free sets of irreps containing 3 \SU{5} chiral families under the assumption that the 
symmetry is broken in stages from \SU{12} $\rightarrow$ \SU{5} $\rightarrow$ SM.  In doing so,
looping over all \SU{12} fermion and Higgs assignments was performed with good fits to the 
input data required.  For this purpose an effective theory approach was used to determine 
the leading order tree-level diagrams for the dim-(4 + n) matrix elements in powers of 
$\epsilon^n$, where $\epsilon$ is the ratio of the \SU{5} to the \SU{12} scale.  The best fit was
obtained by requiring all prefactors to be \order{1}, but the large number of them implies
just a few predictions.  With no discrete flavor symmetry adopted, problems with breaking by 
gravity, domain walls and explanation of its origin can be avoided~\cite{adv1,adv2,adv3}.
On the contrary, with such a large \SU{N} gauge group, a host of heavy fermions is predicted 
which are integrated out at the \SU{5} scale.

The \SU{12} model considered is just one of many possibilities (including other assignments
and larger \SU{N} groups), but its features were among the most attractive found: Each \SU{5}
family supermultiplet can be assigned to a different \SU{12} multiplet in the anomaly-free set.  
In the model considered, only one diagram appears for each matrix element for all 5 mass matrices, 
but some additional contribution is needed to obtain the Georgi-Jarlskog relations.

Among some distracting features we point out the prefactors are determined at the top quark 
scale.  They should be run to the \SU{5} unification scale to test their naturalness.  The 
fit considers only real prefactors, so CP violation is not accommodated, but the fit preferred
$\delta_{CP} = \pi$ over $\delta_{CP} =0$ for the leptonic CP phase.  The complete breaking
of \SU{12} $\rightarrow$ \SU{5} while preserving supersymmetry needs to be worked out in more 
detail and is under further study.

\begin{theacknowledgments}

  One of us (CHA) thanks Kaladi Babu, Rabi Mohapatra, and Barbara Szczerbinska for the kind
invitation to attend and present a talk on this work at the CETUP* Workshop on Neutrino Physics
and Unification in Lead, SD in July 23 - 29, 2012.  He especially appreciated some constructive
suggestions by participants at the Workshop.  He thanks the Fermilab Theoretical Physics
Department for its kind hospitality, where part of this work was carried out.  The work of 
RPF was supported by a fellowship within the Postdoc-Programme of the German Academic Exchange
Service (DAAD).  The work of RPF and TWK was supported by US DOE grant E-FG05-85ER40226.
Fermilab is operated by Fermi Research Alliance, LLC under Contract No. De-AC02-07CH11359 with
the U.S. Department of Energy.

\end{theacknowledgments}

\bibliographystyle{aipproc}

\end{document}

%% file: QuarkMassTermDiagramsTablec.tex
\begin{table}
\renewcommand\boldirrep\relax

\setlength{\arraycolsep}{1pt}
\begin{tabular}{lll}
\hline\\[-4pt]

\multicolumn{3}{l}{\bf Up-Type Quark Mass-Term Diagrams}\\[3pt]
\textbf{Dim 4:} &
	{\bf U33:}&(\irrep{10})\irrepbar{220_3}\ldot(\irrep{5})\higgs{924}\ \ldot(\irrep{10})\irrepbar{220_3}\\
\textbf{Dim 5:} &
	{\bf U23:}&(\irrep{10})\irrepbar{792_2}\ldot(\irrep{1})\higgs{66}\ \ldot\massivefermionpair{(\irrepbar{10})\irrep{220}}{(\irrep{10})\irrepbar{220}}\ldot(\irrep{5})\higgs{924}\ \ldot(\irrep{10})\irrepbar{220_3}\\
      &	{\bf U32:}&(\irrep{10})\irrepbar{220_3}\ldot(\irrep{5})\higgs{924}\ \ldot\massivefermionpair{(\irrep{10})\irrepbar{220}}{(\irrepbar{10})\irrep{220}}\ldot(\irrep{1})\higgs{66}\ \ldot(\irrep{10})\irrepbar{792_2}\\
\textbf{Dim 6:} &
	{\bf U13:}&(\irrep{10})\irrep{495_1}\ldot(\irrep{1})\higgs{220}\ \ldot\massivefermionpair{(\irrepbar{10})\irrep{792}}{(\irrep{10})\irrepbar{792}}\ldot(\irrep{1})\higgs{66}\ \ldot\massivefermionpair{(\irrepbar{10})\irrep{220}}{(\irrep{10})\irrepbar{220}}\ldot(\irrep{5})\higgs{924}\ \ldot(\irrep{10})\irrepbar{220_3}\\
    &	{\bf U31:}&(\irrep{10})\irrepbar{220_3}\ldot(\irrep{5})\higgs{924}\ \ldot\massivefermionpair{(\irrep{10})\irrepbar{220}}{(\irrepbar{10})\irrep{220}}\ldot(\irrep{1})\higgs{66}\ \ldot\massivefermionpair{(\irrep{10})\irrepbar{792}}{(\irrepbar{10})\irrep{792}}\ldot(\irrep{1})\higgs{220}\ \ldot(\irrep{10})\irrep{495_1}\\
    &	{\bf U22:}&(\irrep{10})\irrepbar{792_2}\ldot(\irrep{1})\higgs{66}\ \ldot\massivefermionpair{(\irrepbar{10})\irrep{220}}{(\irrep{10})\irrepbar{220}}\ldot(\irrep{5})\higgs{924}\ \ldot\massivefermionpair{(\irrep{10})\irrepbar{220}}{(\irrepbar{10})\irrep{220}}\ldot(\irrep{1})\higgs{66}\ \ldot(\irrep{10})\irrepbar{792_2}\\
\textbf{Dim 7:} &
	{\bf U12:}&(\irrep{10})\irrep{495_1}\ldot(\irrep{1})\higgs{220}\ \ldot\massivefermionpair{(\irrepbar{10})\irrep{792}}{(\irrep{10})\irrepbar{792}}\ldot(\irrep{1})\higgs{66}\ \ldot\massivefermionpair{(\irrepbar{10})\irrep{220}}{(\irrep{10})\irrepbar{220}}\ldot(\irrep{5})\higgs{924}\ \ldot\massivefermionpair{(\irrep{10})\irrepbar{220}}{(\irrepbar{10})\irrep{220}}\\
    &   &\ldot(\irrep{1})\higgs{66}\ \ldot(\irrep{10})\irrepbar{792_2}\\
    &	{\bf U21:}&(\irrep{10})\irrepbar{792_2}\ldot(\irrep{1})\higgs{66}\ \ldot\massivefermionpair{(\irrepbar{10})\irrep{220}}{(\irrep{10})\irrepbar{220}}\ldot(\irrep{5})\higgs{924}\ \ldot\massivefermionpair{(\irrep{10})\irrepbar{220}}{(\irrepbar{10})\irrep{220}}\ldot(\irrep{1})\higgs{66}\ \ldot\massivefermionpair{(\irrep{10})\irrepbar{792}}{(\irrepbar{10})\irrep{792}}\\
    &   &\ldot(\irrep{1})\higgs{220}\ \ldot(\irrep{10})\irrep{495_1}\\
\textbf{Dim 8:} &
    	{\bf U11:}&(\irrep{10})\irrep{495_1}\ldot(\irrep{1})\higgs{220}\ \ldot\massivefermionpair{(\irrepbar{10})\irrep{792}}{(\irrep{10})\irrepbar{792}}\ldot(\irrep{1})\higgs{66}\ \ldot\massivefermionpair{(\irrepbar{10})\irrep{220}}{(\irrep{10})\irrepbar{220}}\ldot(\irrep{5})\higgs{924}\ \ldot\massivefermionpair{(\irrep{10})\irrepbar{220}}{(\irrepbar{10})\irrep{220}}\\
    &	&\ldot(\irrep{1})\higgs{66}\ \ldot\massivefermionpair{(\irrep{10})\irrepbar{792}}{(\irrepbar{10})\irrep{792}}\ldot(\irrep{1})\higgs{220}\ \ldot(\irrep{10})\irrep{495_1}\\[0.1in]
\multicolumn{3}{l}{\bf Down-Type Quark Mass-Term Diagrams}\\[3pt]
\textbf{Dim 5:} &
	{\bf D32:}&(\irrep{10})\irrepbar{220_3}\ldot(\irrepbar{5})\higgs{924}\ \ldot\massivefermionpair{(\irrepbar{5})\irrepbar{220}}{(\irrep{5})\irrep{220}}\ldot(\irrep{1})\higgs{66}\ \ldot(\irrepbar{5})\irrepbar{792_2}\\
    &	{\bf D33:}&(\irrep{10})\irrepbar{220_3}\ldot(\irrepbar{5})\higgs{924}\ \ldot\massivefermionpair{(\irrepbar{5})\irrepbar{220}}{(\irrep{5})\irrep{220}}\ldot(\irrep{1})\higgs{66}\ \ldot(\irrepbar{5})\irrepbar{792_3}\\
\textbf{Dim 6:} &
	{\bf D31:}&(\irrep{10})\irrepbar{220_3}\ldot(\irrepbar{5})\higgs{924}\ \ldot\massivefermionpair{(\irrepbar{5})\irrepbar{220}}{(\irrep{5})\irrep{220}}\ldot(\irrep{1})\higgs{66}\ \ldot\massivefermionpair{(\irrepbar{5})\irrepbar{792}}{(\irrep{5})\irrep{792}}\ldot(\irrep{1})\higgsbar{220}\ \ldot(\irrepbar{5})\irrepbar{66_1}\\
    &	{\bf D22:}&(\irrep{10})\irrepbar{792_2}\ldot(\irrep{1})\higgs{66}\ \ldot\massivefermionpair{(\irrepbar{10})\irrep{220}}{(\irrep{10})\irrepbar{220}}\ldot(\irrepbar{5})\higgs{924}\ \ldot\massivefermionpair{(\irrepbar{5})\irrepbar{220}}{(\irrep{5})\irrep{220}}\ldot(\irrep{1})\higgs{66}\ \ldot(\irrepbar{5})\irrepbar{792_2}\\
    &	{\bf D23:}&(\irrep{10})\irrepbar{792_2}\ldot(\irrep{1})\higgs{66}\ \ldot\massivefermionpair{(\irrepbar{10})\irrep{220}}{(\irrep{10})\irrepbar{220}}\ldot(\irrepbar{5})\higgs{924}\ \ldot\massivefermionpair{(\irrepbar{5})\irrepbar{220}}{(\irrep{5})\irrep{220}}\ldot(\irrep{1})\higgs{66}\ \ldot(\irrepbar{5})\irrepbar{792_3}\\
\textbf{Dim 7:} &
	{\bf D12:}&(\irrep{10})\irrep{495_1}\ldot(\irrep{1})\higgs{220}\ \ldot\massivefermionpair{(\irrepbar{10})\irrep{792}}{(\irrep{10})\irrepbar{792}}\ldot(\irrep{1})\higgs{66}\ \ldot\massivefermionpair{(\irrepbar{10})\irrep{220}}{(\irrep{10})\irrepbar{220}}\ldot(\irrepbar{5})\higgs{924}\ \ldot\massivefermionpair{(\irrepbar{5})\irrepbar{220}}{(\irrep{5})\irrep{220}}\ldot(\irrep{1})\higgs{66}\ \ldot(\irrepbar{5})\irrepbar{792_2}\\
    &	{\bf D21:}&(\irrep{10})\irrepbar{792_2}\ldot(\irrep{1})\higgs{66}\ \ldot\massivefermionpair{(\irrepbar{10})\irrep{220}}{(\irrep{10})\irrepbar{220}}\ldot(\irrepbar{5})\higgs{924}\ \ldot\massivefermionpair{(\irrepbar{5})\irrepbar{220}}{(\irrep{5})\irrep{220}}\ldot(\irrep{1})\higgs{66}\ \ldot\massivefermionpair{(\irrepbar{5})\irrepbar{792}}{(\irrep{5})\irrep{792}}\ldot(\irrep{1})\higgsbar{220}\ \ldot(\irrepbar{5})\irrepbar{66_1}\\
    &	{\bf D13:}&(\irrep{10})\irrep{495_1}\ldot(\irrep{1})\higgs{220}\ \ldot\massivefermionpair{(\irrepbar{10})\irrep{792}}{(\irrep{10})\irrepbar{792}}\ldot(\irrep{1})\higgs{66}\ \ldot\massivefermionpair{(\irrepbar{10})\irrep{220}}{(\irrep{10})\irrepbar{220}}\ldot(\irrepbar{5})\higgs{924}\ \ldot\massivefermionpair{(\irrepbar{5})\irrepbar{220}}{(\irrep{5})\irrep{220}}\ldot(\irrep{1})\higgs{66}\ \ldot(\irrepbar{5})\irrepbar{792_3}\\
\textbf{Dim 8:} &
	{\bf D11:}&(\irrep{10})\irrep{495_1}\ldot(\irrep{1})\higgs{220}\ \ldot\massivefermionpair{(\irrepbar{10})\irrep{792}}{(\irrep{10})\irrepbar{792}}\ldot(\irrep{1})\higgs{66}\ \ldot\massivefermionpair{(\irrepbar{10})\irrep{220}}{(\irrep{10})\irrepbar{220}}\ldot(\irrepbar{5})\higgs{924}\ \ldot\massivefermionpair{(\irrepbar{5})\irrepbar{220}}{(\irrep{5})\irrep{220}}\\
    &	&\ldot(\irrep{1})\higgs{66}\ \ldot\massivefermionpair{(\irrepbar{5})\irrepbar{792}}{(\irrep{5})\irrep{792}}\ldot(\irrep{1})\higgsbar{220}\ \ldot(\irrepbar{5})\irrepbar{66_1}\\[0.1in]
\hline\\
\end{tabular}
\caption{Leading order up- and down-type quark diagrams for each mass matrix element.}
\label{tab:quarkdiagrams}
\end{table}

%% file: NeutrinoMassTermDiagramsTablec.tex
\begin{table}[t]
\renewcommand\boldirrep\relax

\setlength{\arraycolsep}{1pt}
\begin{tabular}{lll}
\hline\\[-4pt]

\multicolumn{3}{l}{\bf Dirac-Neutrino Mass-Term Diagrams}\\[3pt]
\textbf{Dim 4:} &
	{\bf DN23:} & (\irrepbar{5})\irrepbar{792_2}\ldot(\irrep{5})\higgs{924}\ \ldot(\irrep{1})\irrepbar{12_3}\\
    &	{\bf DN33:} & (\irrepbar{5})\irrepbar{792_3}\ldot(\irrep{5})\higgs{924}\ \ldot(\irrep{1})\irrepbar{12_3}\\
\textbf{Dim 5:} &
	{\bf DN13:} & (\irrepbar{5})\irrepbar{66_1}\ldot(\irrep{1})\higgsbar{220}\ \ldot\massivefermionpair{(\irrep{5})\irrep{792}}{(\irrepbar{5})\irrepbar{792}}\ldot(\irrep{5})\higgs{924}\ \ldot(\irrep{1})\irrepbar{12_3}\\
    &	{\bf DN22:} & (\irrepbar{5})\irrepbar{792_2}\ldot(\irrep{1})\higgs{66}\ \ldot\massivefermionpair{(\irrep{5})\irrep{220}}{(\irrepbar{5})\irrepbar{220}}\ldot(\irrep{5})\higgs{924}\ \ldot(\irrep{1})\irrepbar{220_2}\\
    &	{\bf DN32:} & (\irrepbar{5})\irrepbar{792_3}\ldot(\irrep{1})\higgs{66}\ \ldot\massivefermionpair{(\irrep{5})\irrep{220}}{(\irrepbar{5})\irrepbar{220}}\ \ldot(\irrep{5})\higgs{924}\ \ldot(\irrep{1})\irrepbar{220_2}\\
\textbf{Dim 6:} &
        {\bf DN12:} & (\irrepbar{5})\irrepbar{66_1}\ldot(\irrep{1})\higgsbar{220}\ \ldot\massivefermionpair{(\irrep{5})\irrep{792}}{(\irrepbar{5})\irrepbar{792}}\ldot(\irrep{1})\higgs{66}\ \ldot\massivefermionpair{(\irrep{5})\irrep{220}}{(\irrepbar{5})\irrepbar{220}}\ldot(\irrep{5})\higgs{924}\ \ldot(\irrep{1})\irrepbar{220_2}\\
    &    {\bf DN21:} & (\irrepbar{5})\irrepbar{792_2}\ldot(\irrep{1})\higgs{66}\ \ldot\massivefermionpair{(\irrep{5})\irrep{220}}{(\irrepbar{5})\irrepbar{220}}\ldot(\irrep{5})\higgs{924}\ \ldot\massivefermionpair{(\irrep{1})\irrepbar{220}}{(\irrep{1})\irrep{220}}\ldot(\irrep{1})\higgs{66}\ \ldot(\irrep{1})\irrepbar{792_1}\\
    &	{\bf DN31:} & (\irrepbar{5})\irrepbar{792_3}\ldot(\irrep{1})\higgs{66}\ \ldot\massivefermionpair{(\irrep{5})\irrep{220}}{(\irrepbar{5})\irrepbar{220}}\ldot(\irrep{5})\higgs{924}\ \ldot\massivefermionpair{(\irrep{1})\irrepbar{220}}{(\irrep{1})\irrep{220}}\ldot(\irrep{1})\higgs{66}\ \ldot(\irrep{1})\irrepbar{792_1}\\
\textbf{Dim 7:} &
	{\bf DN11:} & (\irrepbar{5})\irrepbar{66_1}\ldot(\irrep{1})\higgsbar{220}\ \ldot\massivefermionpair{(\irrep{5})\irrep{792}}{(\irrepbar{5})\irrepbar{792}}\ldot(\irrep{1})\higgs{66}\ \ldot\massivefermionpair{(\irrep{5})\irrep{220}}{(\irrepbar{5})\irrepbar{220}}\ldot(\irrep{5})\higgs{924}\ \ldot\massivefermionpair{(\irrep{1})\irrepbar{220}}{(\irrep{1})\irrep{220}}\ldot(\irrep{1})\higgs{66}\ \ldot(\irrep{1})\irrepbar{792_1}\\[0.1in]
\multicolumn{3}{l}{\bf Majorana-Neutrino Mass-Term Diagrams}\\[3pt]
\textbf{Dim 4:} &
	{\bf MN11:} & (\irrep{1})\irrepbar{792_1}\ldot(\irrep{1})\higgsbar{66}\ \ldot(\irrep{1})\irrepbar{792_1}\\
    &	{\bf MN33:} & (\irrep{1})\irrepbar{12_3}\ldot(\irrep{1})\higgs{66}\ \ldot(\irrep1)\irrepbar{12_3}\\
\textbf{Dim 5:} &
        {\bf MN12:} & (\irrep{1})\irrepbar{792_1}\ldot(\irrep{1})\higgsbar{66}\ \ldot\massivefermionpair{(\irrep{1})\irrepbar{792}}{(\irrep{1})\irrep{792}}\ldot(\irrep{1})\higgsbar{66}\ \ldot(\irrep{1})\irrepbar{220_2}\\
    &   {\bf MN21:} & (\irrep{1})\irrepbar{220_2}\ldot(\irrep{1})\higgsbar{66}\ \ldot\massivefermionpair{(\irrep{1})\irrep{792}}{(\irrep{1})\irrepbar{792}}\ldot(\irrep{1})\higgsbar{66}\ \ldot(\irrep{1})\irrepbar{792_1}\\
\textbf{Dim 6:} &
        {\bf MN13:} & (\irrep{1})\irrepbar{792_1}\ldot(\irrep{1})\higgsbar{66}\ \ldot\massivefermionpair{(\irrep{1})\irrepbar{792}}{(\irrep{1})\irrep{792}}\ldot(\irrep{1})\higgsbar{66}\ \ldot\massivefermionpair{(\irrep{1})\irrepbar{220}}{(\irrep{1})\irrep{220}}\ldot(\irrep{1})\higgsbar{66}\ \ldot(\irrep{1})\irrepbar{12_3}\\
    &   {\bf MN31:} & (\irrep{1})\irrepbar{12_3}\ldot(\irrep{1})\higgsbar{66}\ \ldot\massivefermionpair{(\irrep{1})\irrep{220}}{(\irrep{1})\irrepbar{220}}\ldot(\irrep{1})\higgsbar{66}\ \ldot\massivefermionpair{(\irrep{1})\irrep{792}}{(\irrep{1})\irrepbar{792}}\ldot(\irrep{1})\higgsbar{66}\ \ldot(\irrep{1})\irrepbar{792_1}\\
    &   {\bf MN22:} & (\irrep{1})\irrepbar{220_2}\ldot(\irrep{1})\higgsbar{66}\ \ldot\massivefermionpair{(\irrep{1})\irrep{792}}{(\irrep{1})\irrepbar{792}}\ldot(\irrep{1})\higgsbar{66}\ \ldot\massivefermionpair{(\irrep{1})\irrepbar{792}}{(\irrep{1})\irrep{792}}\ldot(\irrep{1})\higgsbar{66}\ \ldot(\irrep{1})\irrepbar{220_2}\\
\textbf{Dim 7:} &
        {\bf MN23:} & (\irrep{1})\irrepbar{220_2}\ldot(\irrep{1})\higgsbar{66}\ \ldot\massivefermionpair{(\irrep{1})\irrep{792}}{(\irrep{1})\irrepbar{792}}\ldot(\irrep{1})\higgsbar{66}\ \ldot\massivefermionpair{(\irrep{1})\irrepbar{792}}{(\irrep{1})\irrep{792}}\ldot(\irrep{1})\higgsbar{66}\ \ldot\massivefermionpair{(\irrep{1})\irrepbar{220}}{(\irrep{1})\irrep{220}}\ldot(\irrep{1})\higgsbar{66}\ \ldot(\irrep{1})\irrepbar{12_3}\\
    &    {\bf MN32:} & (\irrep{1})\irrepbar{12_3}\ldot(\irrep{1})\higgsbar{66}\ \ldot\massivefermionpair{(\irrep{1})\irrep{220}}{(\irrep{1})\irrepbar{220}}\ldot(\irrep{1})\higgsbar{66}\ \ldot\massivefermionpair{(\irrep{1})\irrep{792}}{(\irrep{1})\irrepbar{792}}\ldot(\irrep{1})\higgsbar{66}\ \ldot\massivefermionpair{(\irrep{1})\irrepbar{792}}{(\irrep{1})\irrep{792}}\ldot(\irrep{1})\higgsbar{66}\ \ldot(\irrep{1})\irrepbar{220_2}\\[0.1in]
\hline\\
\end{tabular}
\caption{Leading order Dirac and Majorana neutrino diagrams for each mass matrix element.}
\label{tab:neutrinodiagrams}
\end{table}